\documentclass[conference]{IEEEtran}
\usepackage{biblatex}
\usepackage{amsmath,amssymb,amsfonts}
\usepackage{algorithmic}
\usepackage{graphicx}
\usepackage{textcomp}
\usepackage{xcolor}
\usepackage{fancyhdr}
\usepackage{bigstrut}
\usepackage{multicol}
\usepackage{multirow}
\usepackage{colortbl}
\usepackage{float}
\usepackage{hyperref}
\usepackage{url}
\usepackage{setspace}
\def\BibTeX{{\rm B\kern-.05em{\sc i\kern-.025em b}\kern-.08em
    T\kern-.1667em\lower.7ex\hbox{E}\kern-.125emX}}
    
\begin{document}
 
\title{ Creating Android Malware Knowledge Graph Based on a Malware Ontology}

\author{
  Ahmed Sabbah $^*$\\
  \textit{Computer Science Dept.} \\
  \textit{Birzeit University}\\
  \textit{Ramallah, Palestine}\\
  \textit{asabah@birzeit.edu}
 
  \and
 Mohammed Kharma \\
 \textit{Computer Science Dept.} \\
 \textit{Birzeit University}\\
 \textit{Ramallah, Palestine}\\
 \textit{mkharmah@birzeit.edu}\\

 \and Mustafa Jarrar \\
 \textit{Computer Science Dept.} \\
 \textit{Birzeit University}\\
 \textit{Ramallah, Palestine}\\
 \textit{mjarrar@birzeit.edu}

 }



\maketitle
\thispagestyle{fancy}

\begin{abstract}
As mobile and smart connectivity continue to grow, malware presents a permanently evolving threat to different types of critical domains such as health, logistics, banking, and community segments. Different types of malware have dynamic behaviors and complicated characteristics that are shared among members of the same malware family. Malware threat intelligence reports play a crucial role in describing and documenting the detected malware, providing a wealth of information regarding its attributes, patterns, and behaviors. There is a large amount of intelligent threat information regarding malware. The ontology allows the systematic organization and categorization of this information to ensure consistency in representing concepts and entities across various sources. In this study, we reviewed and extended an existing malware ontology to cover Android malware. Our extended ontology is called AndMalOnt. It consisted of 13 new classes, 16 object properties, and 31 data properties. Second, we created an Android malware knowledge graph by extracting reports from the MalwareBazaar repository and representing them in AndMalOnt. This involved generating a knowledge graph that encompasses over 2600 malware samples. Our ontology, knowledge graph, and source code are all open-source and accessible via GitHub: \href{https://github.com/asabbah44/MalewareOnto}{asabbah44/MalewareOnto} 

\end{abstract}

\section{\textbf{Introduction}}
Mobile malware is an infinite challenge for researchers and industry since the competition between malware authors and defenders will not stop \cite{sabbah2022android}. Two types of operating systems dominate the mobile device market: Android and iOS. Android is open source and holds a market share of approximately 70.79\%, which makes Android malware detection an important area of research due to the increasing number of mobile malware attacks \cite{MobileOp39:online}. Android malware detection can be performed using three main approaches: static, dynamic, and hybrid, which combine both static and dynamic techniques \cite{Alzubaidi2021}. Analysis approaches utilize different methods to extract semantic information and analyze data from various sources \cite{zhang2019efficient}. These methods enable researchers to gain insights into the behavior and characteristics of malware, leading to better detection and understanding of potential threats. \cite{bai2021n}. There are two main objectives for malware analysis. The first is to detect and prevent malware, and the second is to share information about this malware. Multiple sources and websites publish this information for research, industry, and to educate people. VirusTotal is an online service that offers free file and URL scanning. It analyzes submitted items using multiple antivirus engines reaching 70 for malware and suspicious activities. Then, VirusTotal provides a comprehensive report with the detection ratio, names of antivirus engines flagging the file, and additional details \cite{VirusTot96:online}. Additionally, Malware-Bazaar is an online platform that serves as a repository for various malware samples and related information. It allows security researchers and analysts to upload, share, and access malware samples. Malware-Bazaar provides a central database where researchers can contribute and collaborate, allowing the exchange of knowledge and insights about emerging threats \cite{MalwareB41:online}. This information needs to be present in the structure method to construct the knowledge base for the malware domain to use for detection, effective retrieval, and analysis.
The ontology is used in the malware domain for two purposes. Firstly, it plays a crucial role in the detection of malware based on its behavior, as proposed by various studies \cite{gregio2014ontology, navarro2018leveraging, xia2017malware, chowdhury2022capturing}. Secondly, ontology is utilized in the domains of threat intelligence and information security to organize and categorize extensive amounts of threat intelligence data \cite{ding2019ontology, rastogi2020malont, christian2021ontology, mundie2013ontology}.
Based on a lot of threat intelligence information provided in malware, an ontology facilitates the integration of information from various reports, it enables the representation of data from different sources and the reasoning about this data. 
In this paper, we intend to represent the unstructured data for Android malware in the comprehensive knowledge graph. We build the Android Malware Ontology(AndMalOnt) using  Ontology Web Language (OWL). Our AndMalOnt ontology extended available malware ontology named MalOnt2.0 \cite{christian2021ontology}. We reviewed MalOnt2.0 and adopted the classes, object properties, and data properties in AndMalOnt. Additionally, we defined 13 new classes, 16 object properties, and 31 data properties in our ontology. Malware-Bazaar \Cite{MalwareB41:online} was used as a use case to evaluate AndMalOnt by generating a knowledge graph for more than 2600 malware samples.

The rest of this study is structured as follows: Section 2 proposes background about Android malware and threat intelligence. Section 3 presents related work 
Section 4 presents our methodology to create AndMalOnt ontology. Section 5 presents use cases to generate an Android malware knowledge graph based on AndMalOnt. Finally, section 6 presents the conclusion and future work.       
\section{\textbf{Background}}

\subsection{Android Malware}
Malware is a  term derived from \textbf{mal}icious soft\textbf{ware}. Mobile malware aims to gain access to a device with the intent to steal data, harm it, annoy the user, etc. \cite{felt2011survey}. Android malware apps can be installed from official stores, third parties, or using social engineering strategies \cite{Alzubaidi2021}, to gain unauthorized access and use root privileges without the user's permission \cite{felt2011survey}.
Android malware comes in a wide variety of types, and by identifying similar characteristics or behaviors, it is possible to classify them into families. Banking malware, Trojans, Spyware, and Ransomware are some typical instances of Android malware families. 

\subsection{Cyber threat intelligence}
Cyber threat intelligence (CTI) is the process of gathering, analyzing, and acting upon cyber-security data \cite{CyberThr22:online}. CTI aims to provide valuable insights into potential cyber-attacks, it can contain information about the time and location of an attack, malware type used and its hash, the platforms affected, or weakness points to an attack. Additionally,  it includes information about indicators of compromise (IOCs) such as IP addresses, and attack vectors such as phishing emails \cite{gronberg2019ontology}. Sharing CTI information can assist organizations in enhancing their cyber defenses through collaboration, obtaining a better understanding of the threat landscape, and coordinating responses to new threats to mitigate their impact \cite{zheng2015cyber}.
The main components of cyber threat intelligence include: 
\subsubsection{Threat actors}
Threat actors, also known as adversaries, are individuals or organizations responsible for cyber security incidents. Motivations may be political, religious, monetary, or personal, among others. People with limited technical knowledge can use pre-made exploits that are easily accessible online, while experts can discover and exploit zero-day vulnerabilities \cite{gronberg2019ontology}.

\subsubsection{cyber threat intelligence model}
\begin{figure}
    \centering
    \includegraphics[scale=0.4]{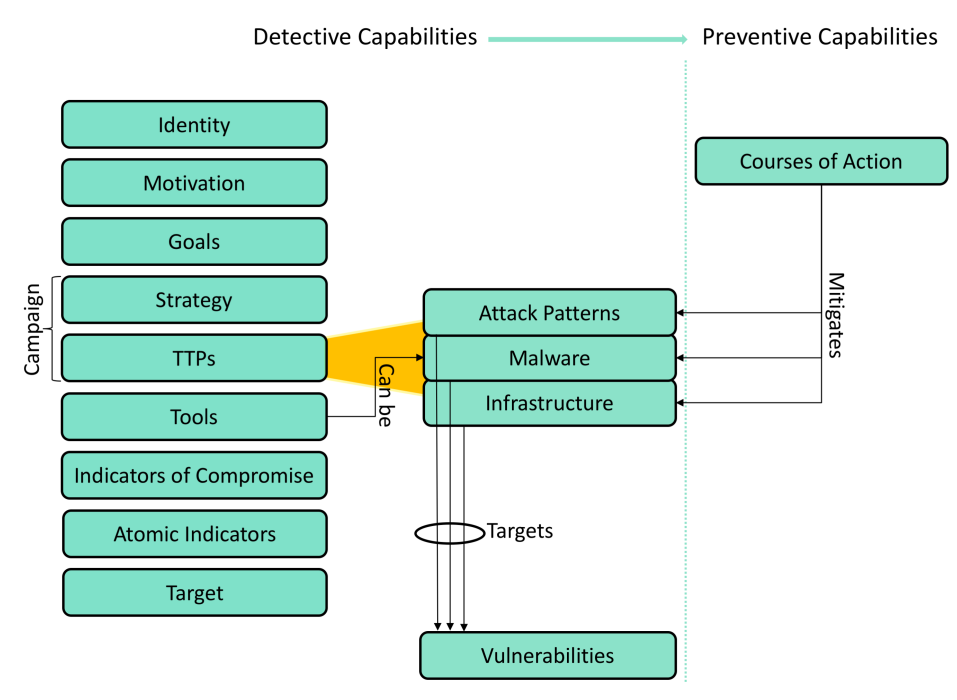}
    \caption{CTI model \cite{mavroeidis2017cyber}}
    \label{fig:CTImodel}
\end{figure}

More than one model was used to represent information regarding threats. For example, the diamond model of intrusion analysis. The cyber-threat intelligence (CTI) model identifies the types of information required for advanced threat intelligence and attack attribution. In addition, it differentiates between the information needed for the prevention and detection of attacks \cite{mavroeidis2017cyber}. The CTI model shows in figure \ref{fig:CTImodel}  

\subsubsection{Taxonomies and information sharing standards}
 MITRE is known for its contributions to the development of cybersecurity frameworks and standards, such as the Common Vulnerabilities and Exposures (CVE) system and the Common Attack Pattern Enumeration and Classification (CAPEC) framework, which describe threat actor behavior. They also maintain the MITRE ATT\&CK framework, which is a knowledge base for adversary tactics, techniques, and procedures (TTPs) used in cyber attacks. MITRE's work in cyber-security aims to improve the understanding of threats, develop effective defenses, and promote information sharing and collaboration among organizations \cite{gronberg2019ontology}. STIX is a representation language that is expressive, flexible, and extensible. It contains various cyber threat information details, such as cyber observables, procedures, incidents, adversary tactics, indicators, techniques, exploit targets, courses of action, cyber attack campaigns, and threat actors \cite{mavroeidis2017cyber}.

Based on the previous information, despite experts' best efforts, context and transparency are lacking when it comes to sharing CTI about malware threats. Data-driven threat intelligence is insufficient for analysts. Existing standards, such as STIX and Trusted Automated eXchange of Indicator Information (TAXII), still ignore important information such as the means to store properties such as malware type (dropper, trojan, etc.), the attacker’s location, and other structured information\cite{rastogi2020malont}.

\section{\textbf{Related work}}
Ontologies are used to represent agreed domain semantics \cite{jarrar2009ontology} and to enable the reusability of such semantics \cite{JM02b}. In cybersecurity, ontology has been used to describe malware knowledge - classes and individuals, and then a reasoner to answer queries about cyber threat data \cite{gronberg2019ontology}. Ontologies were also used for the detection of malware based on its behavior.

\citeauthor{mundie2013ontology} \cite{mundie2013ontology} developed an ontology-based malware analysis to address the lack of a shared vocabulary and concepts in the domain. The data in this study were collected from practitioners, open-source literature, and textbooks. The ontology comprises approximately 270 classes and provides a scientific approach to malware analysis and sharing of information in the field of information security. \citeauthor{gregio2014ontology} \cite{gregio2014ontology} proposed an ontology for malware based on their behavior to identify unknown malware samples and distinguish malicious from benign software. The proposed ontology provides a framework for classifying and representing suspicious behaviors. The authors used available information security ontologies, such as Swimmer, which provides a classification scheme for malware, as well as MAEC and MAL frameworks, that offer vocabulary and taxonomies for describing malware. The same approach was used by \citeauthor{xia2017malware} \cite{xia2017malware} that proposed a malware detection method based on ontology. This method focuses on the behavior of malicious code and creates a knowledge representation of malware behavior from various perspectives. The concepts are divided into two main classes: malware knowledge and system components. To represent the behavior of a malware family, this method utilizes the common behavior of individuals. An ontology reasoning mechanism was then employed to detect unknown malware samples. Additionally, an ontology-based framework was proposed to model the interactions between application and system aspects.

In this approach, machine learning techniques were employed to analyze complex networks and identify shared features among various malware samples \cite{navarro2018leveraging}. To deep insight into malware behavior, 
\citeauthor{chowdhury2022capturing} \cite{chowdhury2022capturing} proposes an ontology-based framework for capturing malware behavior and compares it with two other ontologies, MALOnt and Swimmer. The suggested ontology captures the modifications of metamorphic malware API call sequences and provides a deep understanding of various malware types and their behavior. This paper also provides an overview of advanced techniques used by malware writers to avoid detection and explains how ontology can help in conceptualizing the domain knowledge of malicious behavior. The ontology includes two sub-categories: Malware Families and malware code structures. Malware families have 14 subclasses.

All the above studies proposed the use of ontology for malware detection by representing the relationship between malware and its behavior. The second uses ontology to provide information about the already detected malware and to share information about the results of the first direction from different sources. 
\citeauthor{ding2019ontology} \cite{ding2019ontology} 
created a comprehensive knowledge base that can effectively store and represent behavioral information regarding individual malware instances and families. The primary aim is to assist users in malware analysis and detection using an ontology to describe malware behavior and establish a structured knowledge framework. In addition, ontology reasoning techniques have been used to identify families of unknown 
malware. Therefore, studies have focused on building a knowledge graph to share information using ontologies. 
\citeauthor{rastogi2020malont} \cite{rastogi2020malont} introduced MALOnt, an open-source malware ontology designed to collect and organize malware threat intelligence from various online sources. MALOnt contains a wide range of concepts related to malware, including its characteristics, attack details, and attacker information. Moreover, it depends on the previous ontology to collect information \cite{costa2016insider,iannacone2015developing, swimmer2008towards}. MALOnt allows for the creation of a knowledge graph by populating specific instances within the ontology. This paper highlights the importance of the structured extraction of information and the generation of knowledge graphs for analyzing, detecting, classifying, and cyber threats caused by malware. This ontology defines 68 classes, 31 properties, and 13 properties, respectively. To generate a knowledge graph from this ontology, security reports and Name Entity Recognition(NER) are used to extract individuals. The annotation process uses tools, such as the Brat Rapid Annotation Tool and INCEpTION, which help in labeling and organizing information from threat intelligence reports. These tools are used to label specific text segments in the reports as MALOnt classes, such as identifying "PowerPoint file" as the class "Software" and "installs malicious code" as the class "Vulnerability.” The relationships between these classes are represented by arrows, which indicate semantic connections.  
\citeauthor{christian2021ontology} \cite{christian2021ontology} proposed MalONT2.0 which is a significant improvement over the prior version \cite{rastogi2020malont}. This study used NER to annotate cyber-threat intelligence (CTI) reports on Android malware attacks. Each annotation in the reports was instantiated into classes and shared relationships with the other instances. These classes and relations are defined and described in MalONT2.0. and share similarities with the STIX2.1 framework where appropriate. The instances representing the annotated information were stored in RDF (Resource Description Framework) triples. Each triple consists of an entity, relation, and entity tail. This structure captures the components of the CTI reports and forms a basis for the knowledge graph. The results of representing 25 CTI reports written between 2011 -2021, and using (NER) to annotate 1,100 entities and 2,300 relations.

\section{\textbf{Extending MalOnt2.0 }}
 The main objective of this study is to create an ontology-based knowledge graph for threat intelligence about Android malware. To do this, we adopted an existing ontology called MalOnt2.0 \cite{christian2021ontology} that provides knowledge about malware. We extended this ontology with concepts and relations to describe Android malwares that are not covered in MalOnt2.0. Our extension MalOnt2.0 is provided as a separate ontology, which is an important design criteria in ontology engine engineering \cite{J05a, J05}. To evaluate our AndMalOnt, we adopted a large repository of malware called malware Bazaar website \footnote{https://bazaar.abuse.ch/}. We extracted a knowledge graph of about 2600 Android malware reports and represented them in RDF using our ontology.  

 \subsection{\textbf{The MalOnt2.0 ontology}}
 
\begin{figure}[h]
    \centering
    \includegraphics[scale=0.4]{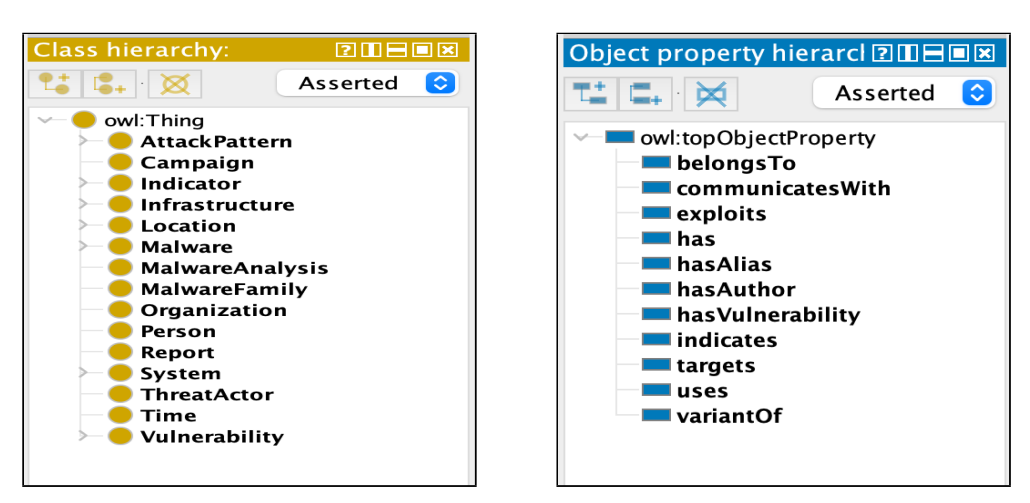}
    \caption{Main Classes (left), properties (right) \cite{christian2021ontology} }
    \label{fig:Malont2}
\end{figure}

 In this paper, we focus on the ontology named MalOnt2.0 \cite{christian2021ontology}, which used CTI reports and STIX 2 to create a malware ontology. The main class and properties are shown in figure \ref{fig:Malont2}. 

The definition of classes in MalOnt2.0 is as the following: 
 \begin{itemize}
     \item \textbf{Attack Pattern}: The method of achieving an attack that exploits specific vulnerabilities in a certain environment. It uses Tactics, Techniques, and Procedures (TTPs) of the adversaries' attempt to compromise targets. Example CAPAC[604] = Wi-Fi Jamming.

     \item \textbf{Campaign :} "A grouping of adversarial behaviors that describes a set of malicious activities or attacks (sometimes called waves) that occur over a period of time against a specific set of targets" (Adapted from STIX 2.1). 

     \item \textbf{Indicator :} Indicators of compromise are distinguishable artifacts in a computer system that indicate malicious or suspicious behavior. For example, IP address, email address, port, and other subclass defined under the Indicator class.

      \item \textbf{Infrastructure :} Describes any software, systems, services, and any physical or virtual resources planned to support some objective, whether in attack or defense (eg: Command and controller attack) (Adapted from STIX 2.1).

      \item \textbf{Location :} The geographic location of a place.

     \item  \textbf{Malware: } Malicious software intended to violate the integrity, availability, or confidentiality of a computer system.

     \item \textbf{Malware analysis: } Malware analysis captures the VirusTotal extracted static or dynamic analysis performed on a malware instance or family (from STIX 2.1).

     \item \textbf{Malware Family : } Group of malware with common properties. 

     \item \textbf{Organization : } An organization that is either responsible for the attack or has been attacked by an adversary.

    \item \textbf{Person :} Attacker or person Attacked.

    \item  \textbf{Report :} "Reports are collections of threat intelligence focused on one or more topics, such as a description of a threat actor, malware, or attack technique, including context and related details (from STIX 2.1)". 

    \item  \textbf{System :} Describes the hardware and software specifications (both are defined as subclasses).

    \item  \textbf{Threat Actors :} are actual individuals, groups, or organizations believed to be operating with malicious intent?

    \item  \textbf{Time :} The time of an event in different formats can be absolute or relative.

     \item \textbf{Vulnerability :} A vulnerability refers to a defect or weakness in the requirements, designs, or implementations of computational logic, such as code in software or firmware in hardware components. These vulnerabilities can be exploited to compromise the confidentiality, integrity, or availability (CIA) of the affected system, which is a potential risk to its overall security and functionality. (STIX 2.1)
 \end{itemize}

 \subsection{\textbf{Our AndMalOnt ontology Module}}

\begin{figure}[h]
    \centering
    \includegraphics[width=8cm,height=3.5cm] {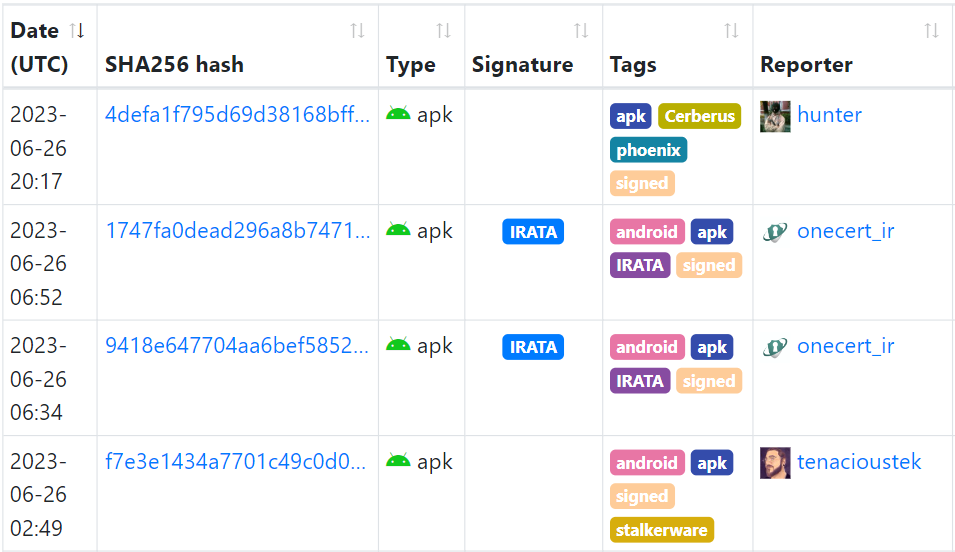}
    \caption{Malware bazaar website}
    \label{fig:MalBazzar}
\end{figure}

We undertake a thorough review of various sources related to malware reports and ontologies. This includes Malont2.0, which is a well-known ontology for malware analysis. We also explore malware repositories websites such as Malware Bazaar \cite{MalwareB41:online}, and Virus-Total \cite{VirusTot96:online}, and analyze thread intelligence reports provided by trusted sources like Kaspersky \footnote{https://kaspersky.com/} and Avast \footnote{https://www.avast.com/}. By examining these diverse sources, we aim to collect up-to-date concepts about Android malware. Figure \ref{fig:MalBazzar} presents information from Malware-Bazaar, which shows the essential information about Android malware such as the malware family that appears in the signature column. 

From these information sources, we observe that in addition to the classes and properties that were defined in MalOnt2.0, we need to define the following classes:    
\begin{itemize}
\vspace{0.5em}
\item \textbf{Hashing: } Malont2.0 defined five types of hashing (\textit{MD5, SHA-1, SHA-256, SSDeep, vHash}), and some of these hashing algorithms are used to share this hash that indicates the file is malware. This type is used by antivirus software to detect malware (e.g., SHA-256) \cite{HashAlgo79:online}. Other types of hashing are used to provide a comparison or similarity measure between hashing of malware to detect the malware's family, such as TLSH. Thus, we defined new hashing classes in AndMalOnt, which are considered a subclass of \textit{HASH} class in MalOnt 2.0. These classes: \textit{IMPHASH, TLSH, TELIFHASH, GIMHASH}. Additionally, to define \textit{SHA2, and SHA3} since the difference between instances of these types is the size of the bits, we added a new enumerate class \textit{HashDigestSize} that is a subclass of \textit{HASH} in MalOnt2.0. Figure \ref{fig:HASH2} presents all hash classes.

\begin{figure}[h]
    \centering
    \includegraphics[scale=0.5]{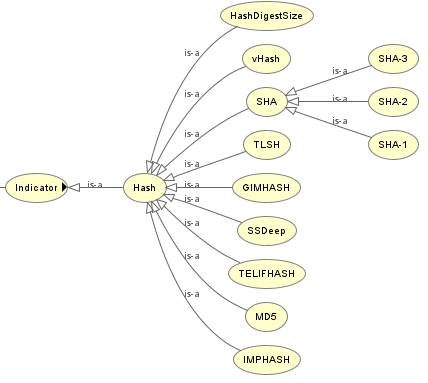}
    \caption{AndMalOnt Hash classes }
    \label{fig:HASH2}
\end{figure}

\item \textbf{File: } The \textit{File} class represents the fundamental entity that contains information about a specific file associated with Android malware. This class is designed to capture attributes, properties, and relationships related to the file itself. The \textit{File} class can include properties such as file name, file size, file type, and file path. These properties provide basic information about the Android malware file. Additionally, the "File" class can be connected to other relevant classes in ontology, such as "Malware" or "Certificate," to establish relationships and associations between the file and other entities.

\item \textbf{Malware reporter :} is the entity who first reported the file as malware, This entity could be a person, a company, or even an anonymous source. The \textit{MalwareReporter} class is defined as a new ontology class that encompasses this specific role and its associated information. It allows for the identification and tracking of the reporter of the infected file.

\item \textbf{Publisher: } is defined within the ontology to describe the entity or organization responsible for developing and publishing malware. The \textit{AppPublisher} class allows for the identification and categorization of different actors involved in the creation and publish the malware. This information can be valuable for understanding and tracking the sources, characteristics, and motivations of different malware.

\item  \textbf{Certificate: }  is the code signing certificate that the publisher of the application use in order to sign the application before publishing it. Each publisher might have one code signing certificate or more. We defined a new class (\textit{Certificate}) that contains information about the code sign such as "Thumbprint algorithm". 


\item \textbf{Tag :} The \textit{Tag} class represents individual descriptors or labels associated with malware samples. Each malware sample can be associated with one or more tags, providing additional information and context about the characteristics, behavior, or attributes of the malware. The \textit{Tag} class is connected to the \textit{Malware} class through a relation. This relation can be represented using a property, such as \textit{hasTag} that indicates the association between malware and its corresponding tags. 

 \item \textbf{Vendor intelligence: } More than 70 anti-viruses shared information about specific malware. Some of them claim that the file is malware and some of them claim in benign. Figure \ref{fig:vendors} shows one source reported the file is not malware and others reported it as malware and provided the link for more information.
 
    \begin{figure}[h]
        \centering
        \includegraphics[scale=0.28]{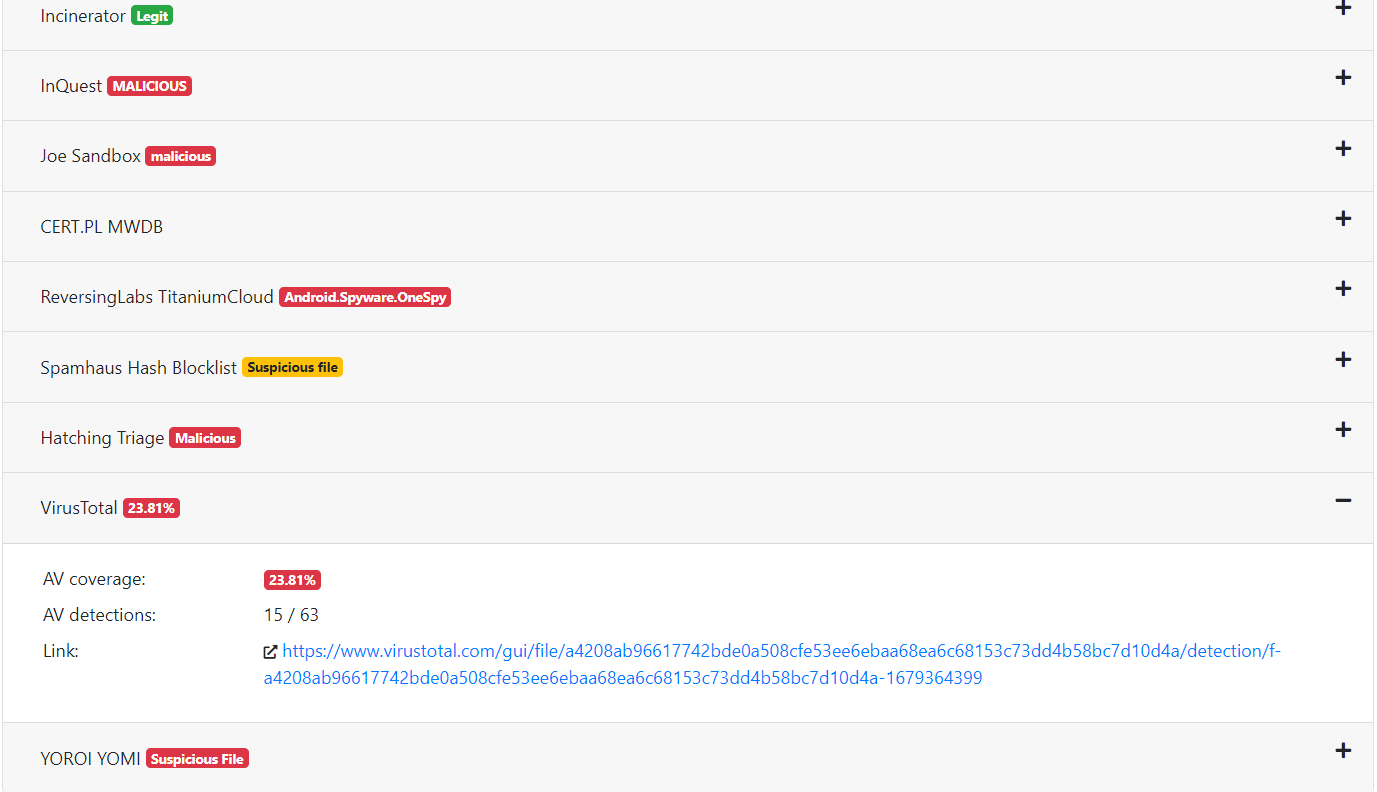}
        \caption{Malware bazaar vendor threat intelligence example}
        \label{fig:vendors}
    \end{figure}
The vendors are defined as a new class, which indicates specific vendor threat intelligence report information.

\item  \textbf{YARA rule} YARA is a popular tool and language for writing rules to identify patterns or signatures associated with known malware. YARA rules are defined as a subclass of "MalwareAnalysis," it can capture the specific properties of YARA rules within the context of malware detection and classification \cite{WritingY39:online}. YARA has properties like "name," "author," "description," and "reference."
Figure \ref{fig:IntRule} shows the final AndMalOnt ontology.

\begin{figure}[h]
    \centering
    \includegraphics[scale=0.1]{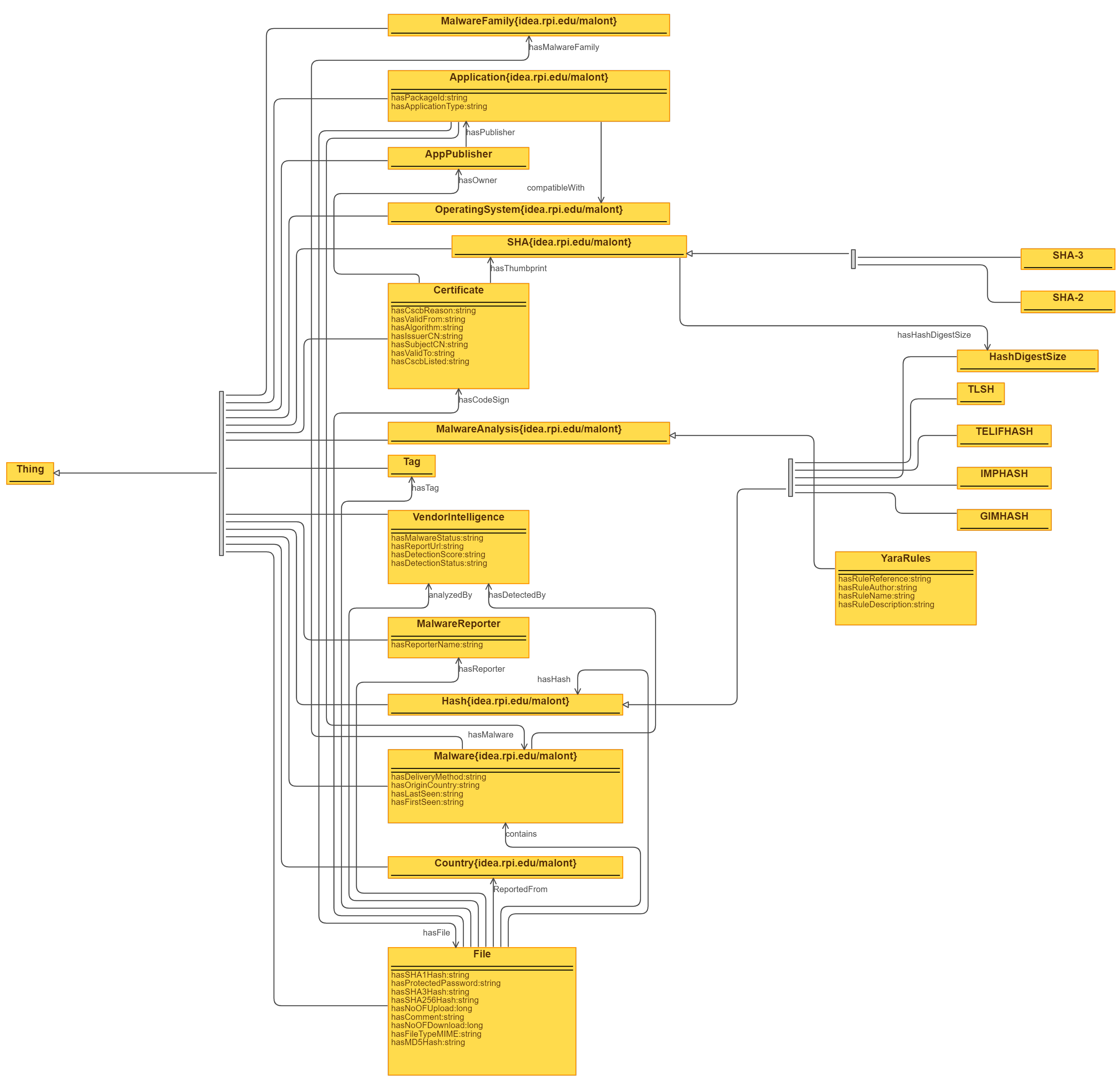}
    \caption{Our AndMalOnt ontology.}
    \label{fig:IntRule}
\end{figure}

\end{itemize}

\subsection{AndMalOnt object properties}
\begin{figure}[h]
    \centering
    \includegraphics [scale=0.7]{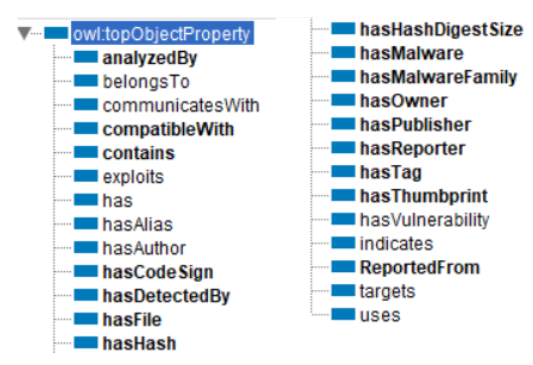}
    \caption{AndMalOnt object properties highlighted in bold text }
    \label{fig:ObjProp}
\end{figure}

In the AndMalOnt ontology, we have defined a total of 16 object properties to show relationships and connections between different classes. These object properties play an important role in capturing the complex associations within the Android malware domain. From these object properties the relationships between malware and its publisher, the connection between malware and their corresponding families. Figure \ref{fig:ObjProp} presents the AndMalOnt object properties.

\subsection{AndMalOnt data properties} 
\begin{figure}[ht]
    \centering
    \includegraphics [scale=0.7]{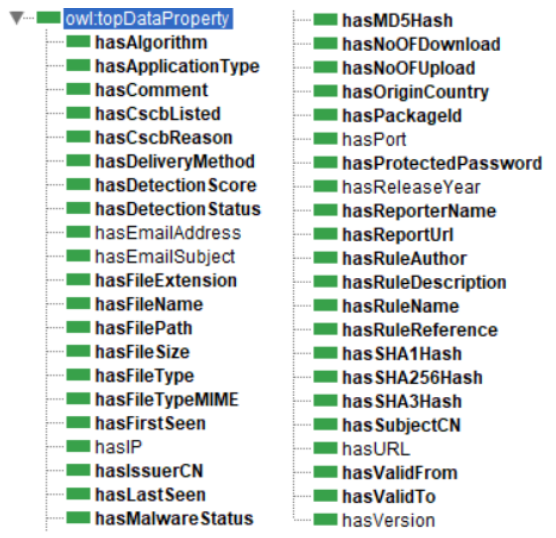}
    \caption{AndMalOnt data properties shown in bold text  }
    \label{fig:DataProp}
\end{figure}

In addition to the object properties, the AndMalOnt ontology contains different data properties to capture specific attributes and characteristics of Android malware. These data properties provide valuable information that enriches the understanding of malware samples and related entities. Examples of commonly used data properties in AndMalOnt included properties that capture the file size, first seen, and last seen date. The final data properties in  AndMalOnt were 31 properties shown in figure \ref{fig:DataProp}

\section{Android malware knowledge graph} 
To evaluate the effectiveness of AndMalOnt, we developed automation tools using the Apache Jena Java library \cite{ApacheJe13:online}. These tools enabled us to generate instances for 2680 Android malware reports using the Malware-Bazaar API described in Table \ref{tab:MalSample}.

\begin{table}[ht]
  \centering
    \begin{tabular}{|l|r|l|r|}
    \hline
   \textbf{Signature} & \multicolumn{1}{l|}{\textbf{Count}} & \textbf{Signature} & \multicolumn{1}{l|}{\textbf{Count}} \\
    \hline
     AbereBot & 8     &  Hydra & 120 \\
    \hline
     AgentSmith & 3     &  IRATA & 118 \\
    \hline
     Alien & 92    &  Joker & 197 \\
    \hline
     Alienbot & 1     &  Kimsuky & 1 \\
    \hline
     Anubis & 21    &  MaliBot & 2 \\
    \hline
     Bahamut & 7     &  Metasploit & 1 \\
    \hline
     BankBot & 11    &  MoqHao & 2 \\
    \hline
     BARAT & 1     &  Multiverze & 1 \\
    \hline
     BasBanke & 1     &  n/a  & 1622 \\
    \hline
     BrasDex & 2     &  Octo & 4 \\
    \hline
     BRATA & 31    &  Pegasus & 3 \\
    \hline
     Cerber & 2     &  PixStealer & 1 \\
    \hline
     Cerberus & 217   &  QNodeService & 1 \\
    \hline
     DoNot & 2     &  Raddex & 1 \\
    \hline
     Dracarys & 1     &  SARA & 2 \\
    \hline
     Ermac & 16    &  SharkBot & 26 \\
    \hline
     Escobar & 1     &  SideWinder & 7 \\
    \hline
     Exobot & 3     &  SMSWorm & 1 \\
    \hline
     FakeCop & 66    &  Sova & 12 \\
    \hline
     FluBot & 27    &  SpyNote & 26 \\
    \hline
     FluHorse & 1     &  TeaBot & 13 \\
    \hline
     Godfather & 4     &  Wroba & 2 \\
    \hline
     Harly & 2     &  XLoader & 2 \\
    \hline
     Heodo & 1     & \cellcolor[rgb]{ .851,  .851,  .851}\textbf{Total} & \cellcolor[rgb]{ .851,  .851,  .851}\textbf{2686} \\
    \hline
    \end{tabular}%
    \vspace{1em}
      \caption{Android malware families based on Malware-Bazaar }
  \label{tab:MalSample}%
\end{table}%

\begin{figure}[h]
    \centering
    \includegraphics[scale=0.30]{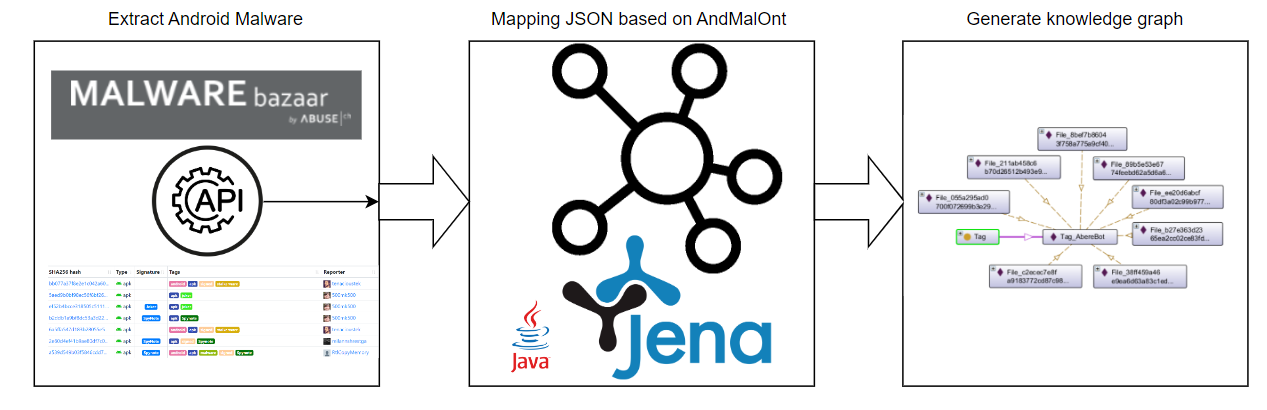}
    \caption{Extracting Android malware to generate KG}
    \label{fig:pipeline}
\end{figure}

 The pipeline used to extract the data from Malware-Bazaar and generate individuals based on the relations and concepts defined in AndMalOnt is presented in figure \ref{fig:pipeline}.

This knowledge graph is a valuable resource for analyzing and understanding the characteristics and behaviors of Android malware. Additionally, to facilitate querying and retrieval of information from the generated knowledge graph, we used SPARQL to construct complex queries to extract specific information, explore relationships between entities, and gain insights into the characteristics and attributes of Android malware instances.

\subsubsection{Use case 1} 
\begin{figure}[ht]
    \centering
    \includegraphics[scale=0.3]{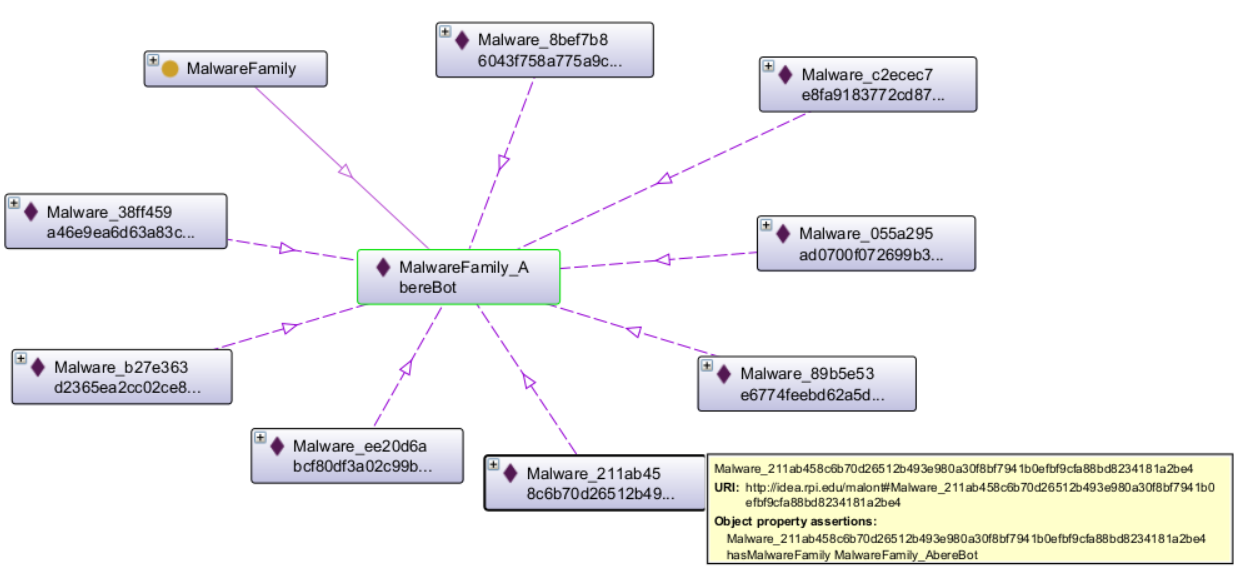}
    \caption{Instances of \textit{"bereBot"} Android Malware family }
    \label{fig:KG-bereBot}
\end{figure}

Find all malware that belong to \textit{"bereBot"} family shows in figure \ref{fig:KG-bereBot}.

\subsubsection{Use case 2 }
\begin{figure}[h]
    \centering
    \includegraphics[scale=0.45]{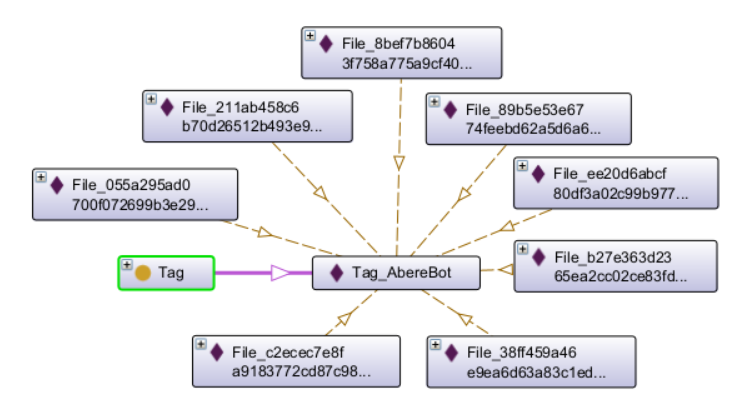}
    \caption{Instances of \textit{"bereBot"} that used as a tag }
    \label{fig:KG-bereBot2}
\end{figure}

For the same malware family \textit{"bereBot"}, figure \ref{fig:KG-bereBot2} presents the malware that uses \textit{"bereBot"} concept as a tag of malware. 

\subsubsection{Use case 3} 
\begin{figure}[h]
    \centering
    \includegraphics [scale=0.3]{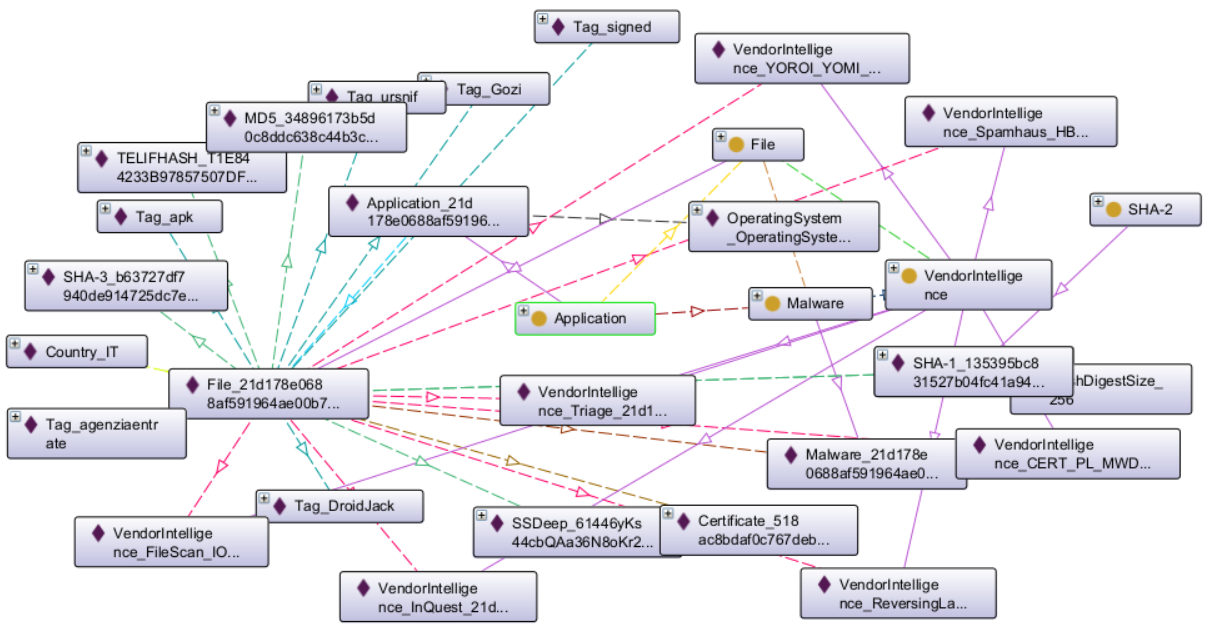}
    \caption{knowledge graph for specific malware.}
    \label{fig:KG-ByHash}
\end{figure}

This part of the knowledge graph was generated based on the hash:

"SHA256 21d178e0688af591964ae00b71263d2e086706017 ebc98d7488d57771144d337". Figure \ref{fig:KG-ByHash} shows the output of the knowledge graph for this specific malware and its relations with other entities.

\subsubsection{Use case 4} 
Discovery of sequence of malware (file that contains more than one malware instance).

\begin{small}
\begin{verbatim}
PREFIX android_malware_ontology: 

<http://secuirty.birzeit.edu/
android_malware_ontology#>

SELECT ?file ?fileName (COUNT(?malwareFamily)
AS ?count)
WHERE { 
?file android_malware_ontology:contains ?malware .
?malware android_malware_ontology:hasMalwareFamily
?malwareFamily .
?file android_malware_ontology:hasFileName ?fileName . 
        }
        GROUP BY ?file ?fileName
        HAVING (COUNT(?malwareFamily) > 1)
\end{verbatim}
\end{small}

\subsubsection{Use case 5 } This query retrieves files with filename, reporter and reporting location.

\begin{small}
\begin{verbatim}
PREFIX android_malware_ontology:
<http://secuirty.birzeit.edu/
android_malware_ontology#>
PREFIX malont: <http://idea.rpi.edu/malont#>

SELECT ?file ?fileName ?reporter ?reportedFrom
WHERE { 
?file android_malware_ontology:ReportedFrom 
?reportedFrom .
?file malont:hasReporter ?reporter .
?file android_malware_ontology:hasFileName 
                ?fileName . 
    }
\end{verbatim}
\end{small}

\subsubsection{Use case 6} To retrieves the top countries that have reported malware based on the AndMalOnt knowledge graph.
\begin{small}
\begin{verbatim}
PREFIX android_malware_ontology:
<http://secuirty.birzeit.edu/
android_malware_ontology#>
PREFIX malont: <http://idea.rpi.edu/malont#>

SELECT ?reportedFrom 
        (COUNT(?reportedFrom) AS ?count)
WHERE { 
?file android_malware_ontology:ReportedFrom
?reportedFrom .
?file malont:hasReporter ?reporter .
    }
    GROUP BY ?reportedFrom
    HAVING (COUNT(?reportedFrom) > 10)
	ORDER BY DESC(?count)
\end{verbatim}
\end{small}

\section{\textbf{Conclusion and future work}}
Malware threat intelligence reports play a critical role in describing and documenting the detected malware. It provides valuable information regarding malware attributes, patterns, and behaviors. The existing malware ontology allows the organization and categorization of this knowledge. Furthermore, the use of ontology enables consistency in representing concepts and entities collected from different sources. In this study, we extended an existing malware ontology to cover Android Malware. Our ontology module is called AndMalOnt.
We also constructed a knowledge graph of Android malware extracted from the MalwareBazaar repository. The knowledge graph consists of 2600 nodes (i.e., individuals) represented in the RDF using the AndMalOnt ontology. Our ontology and knowledge graph are open-source and accessible via GitHub.

In future work, we intend to enrich the AndMalOnt ontology by including additional ontologies that focus on the specific aspects of malware detection. Integrating these complementary ontologies that handle Android malware behavior analysis, code signatures, network traffic, anomaly detection, and evasion techniques will contribute to the continuous improvement and effectiveness of AndMalOnt as a powerful resource for Android malware detection and mitigation.

\printbibliography
\end{document}